\documentclass[twocolumn,prb,aps,superscriptaddress]{revtex4-1}

\let\oldaddcontentsline\addcontentsline
\newcommand{\stoptocentries}{\renewcommand{\addcontentsline}[3]{}}
\newcommand{\starttocentries}{\let\addcontentsline\oldaddcontentsline}

\usepackage[latin9]{inputenc}
\usepackage{mathtools}
\usepackage{amsbsy}
\usepackage{amstext}
\usepackage[unicode=true,pdfusetitle,
 bookmarks=false,
 breaklinks=false,pdfborder={0 0 1},backref=false,colorlinks=false]
 {hyperref}
\hypersetup{
 bookmarksnumbered=false,bookmarksopen=false}
\makeatletter

\@ifundefined{textcolor}{}
{
 \definecolor{BLACK}{gray}{0}
 \definecolor{WHITE}{gray}{1}
 \definecolor{RED}{rgb}{1,0,0}
 \definecolor{GREEN}{rgb}{0,1,0}
 \definecolor{BLUE}{rgb}{0,0,1}
 \definecolor{CYAN}{cmyk}{1,0,0,0}
 \definecolor{MAGENTA}{cmyk}{0,1,0,0}
 \definecolor{YELLOW}{cmyk}{0,0,1,0}
}

\usepackage{amsmath}
\usepackage{graphicx}
\usepackage{amssymb}
\usepackage{txfonts,color}
\makeatother

\begin{document}
\title{Double Quantum Magnetometry at Large Static Magnetic Fields}
\author{C. Munuera-Javaloy}
\affiliation{Department of Physical Chemistry, University of the Basque Country UPV/EHU, Apartado 644, 48080 Bilbao, Spain}
\author{I. Arrazola}
\affiliation{Department of Physical Chemistry, University of the Basque Country UPV/EHU, Apartado 644, 48080 Bilbao, Spain}
\author{E. Solano}
\affiliation{Department of Physical Chemistry, University of the Basque Country UPV/EHU, Apartado 644, 48080 Bilbao, Spain}
\affiliation{IKERBASQUE,  Basque  Foundation  for  Science,  Maria  Diaz  de  Haro  3,  48013  Bilbao,  Spain}
\affiliation{International Center of Quantum Artificial Intelligence for Science and Technology~(QuArtist)
and Department of Physics, Shanghai University, 200444 Shanghai, China}
\author{J. Casanova}
\affiliation{Department of Physical Chemistry, University of the Basque Country UPV/EHU, Apartado 644, 48080 Bilbao, Spain}
\affiliation{IKERBASQUE,  Basque  Foundation  for  Science,  Maria  Diaz  de  Haro  3,  48013  Bilbao,  Spain}

\begin{abstract} 
We present a protocol to achieve double quantum magnetometry at large static magnetic fields. This is a regime where sensitive sample parameters, such as the chemical shift, get enhanced facilitating their characterization. In particular, our method delivers two-tone stroboscopic radiation patterns with modulated Rabi frequencies to achieve larger spectral signals. Furthermore, it  does not introduce inhomogeneous broadening in the sample spectrum preventing signal misinterpretation. Moreover, our protocol is designed to work under realistic conditions such as the presence of moderate microwave power and errors on the radiation fields. Albeit we particularise to nitrogen vacancy centers, our protocol is general, thus applicable to distinct quantum sensors.
\end{abstract}
\maketitle

\stoptocentries
\section{Introduction} The detection of magnetic signals emitted by spin ensembles is central in nanoscale nuclear magnetic resonance (nanoscale NMR)~\cite{Degen17}.  Here, the hyperfine quantum state of a sensor such as the nitrogen vacancy (NV) center in diamond~\cite{Doherty13, Dobrovitski13} gets modified owing to the interaction with a nuclear or electronic target, leading to quantum sensing or quantum detection~\cite{Rondin14, Wu16}. To this end, the NV center quantum state is initialised and measured with a laser field, while it can be readily  controlled with microwave (MW) radiation~\cite{Doherty13, Dobrovitski13}. In addition, the NV center presents quantum coherence, as well as a long decay time of the order of milliseconds, even at room temperature~\cite{Schirhagl14}. These are capacities that make NV centers in diamond ideal candidates for experiments at physiological conditions~\cite{Wu16}. The coupling of the NV center with a target signal is typically conducted trough dynamical decoupling (DD) techniques~\cite{Souza12} that, in addition, are able to remove noisy contributions over the sensor. Typically, continuous and pulsed (or stroboscopic) DD techniques are considered. The former relies on the Hartmann-Hahn resonance condition~\cite{Hartmann62}, while certain stroboscopic DD radiation patterns, as those of the XY family~\cite{Maudsley86, Gullion90, Souza11, Casanova15, Wang17, Lang17, Arrazola18}, offer a superior level of robustness against errors on the control fields. First experiments with NV centers were able to detect  classical electromagnetic fields~\cite{Taylor08}, while individual $^{13}$C nuclear spin emitters embedded in the diamond lattice have been identified with unprecedented resolution ~\cite{Kolkowitz12, Taminiau12, Zhao12, London13, Zopes18, Zopes18bis, Bradley19, Abobeih19}.  In addition, nanoscale NMR of small volume samples of the order of picolitres, as well as of single molecules located external to the diamond lattice, have been achieved with single NV centers~\cite{Mamin13, Staudacher13, Muller14, Lovchinsky16,  Shi15, Aslam17} and with NV ensembles~\cite{DeVience15,Glenn18}.  

Several developments have been carried out to overcome the poor spectral resolution achievable with NV quantum sensors at room temperature. Among them, we have the combination of NV-based quantum detection with the presence of a quantum memory~\cite{Rosskopf17}, or by  synchronising NV measurements with a classical clock~\cite{Schmitt17, Boss17, Glenn18}. These hybrid techniques have allowed the detection of coherent target signals with a frequency resolution of the order of  few hertz with NV centers. Here, it is important to remark that, for all these applications the achievement of an NV-target coupling remains as the key ingredient, and the use of double quantum magnetometry (DQM) schemes~\cite{Reinhard12, Fang13, Mamin14} permits enhanced signal acquisition. Furthermore, DQM techniques do not introduce a magnetic field gradient on the studied sample, which avoids inhomogeneous broadening in the measured spectrum~\cite{Glenn18}.  Especially interesting for nanoscale NMR is the regime of large static magnetic fields~\cite{Aslam17, Casanova18, Casanova19, Arrazola19}. In these conditions, a large chemical shift is present~\cite{Aslam17}. Note this is an important quantity that encodes structural information of the target sample such as the presence of different chemical bonds~\cite{Levitt08}. Furthermore, at large static magnetic fields the relaxation time of sensor gets enhanced~\cite{Reynhardt01}, which provides with better frequency selectivity of the nuclear target~\cite{Rosskopf17}. Consequently, having DQM schemes ready to be displayed at large static magnetic fields is of clear relevance in nanoscale NMR.

In this article, we provide DQM schemes able to conduct sensor-target coupling at large static magnetic fields. In addition, control errors over the sensor get removed owing to the intrinsic DD behaviour carried by our method. Via the use of two-tone stroboscopic MW radiation with modulated field amplitudes, we demonstrate the achievement of enhanced sensor-signal interaction with moderate power in the MW drivings. More importantly, our method removes  inhomogeneous broadenings over the sample avoiding  misinterpretation in the obtained spectrum. Hence, this is an optimal scheme for detecting natural deviations in the resonance frequencies of the target sample, such as those originated from chemical shifts. Our protocol is general, i.e., it can be displayed over different quantum sensors such as NV centers and silicon vacancy centers, while it can be adapted to any stroboscopic DD sequence used in nanoscale NMR.

\section{The method} The Hamiltonian of an NV center in diamond under a  MW driving $ \sqrt{2}B^x_d \cos{(\omega_{\rm MW} t + \varphi)}$ reads
\begin{equation}\label{rotations}
H = DS_z^2 + |\gamma_e| B_z S_z + \sqrt{2}\gamma_e B^x_d \cos{(\omega_{\rm MW} t + \varphi)} S_x. 
\end{equation}
Here $D=(2\pi)\times 2.87$ GHz is the zero field splitting, $|\gamma_e|=(2\pi)\times28.024$ GHz/T,  $B_z$ is the static magnetic field aligned with the NV axis, while  $\omega_{\rm MW}$ and  $\varphi$ are the MW frequency and phase respectively. The spin-1 operators in Eq.~(\ref{rotations}) are $S_z = |1\rangle\langle 1| -  |-1\rangle\langle -1|$, and $S_x = \frac{1}{\sqrt{2}} \big(|1\rangle\langle 0| +  |0\rangle\langle 1| + |-~1\rangle\langle 0| +  |0\rangle\langle -1| \big)$.  The driving term $ \sqrt{2}\gamma_e B^x_d \cos{(\omega_{\rm MW} t + \varphi)} S_x$ in Eq.~(\ref{rotations}) leads to rotations in the NV hyperfine spin states, while an inspection to the $S_x$ operator indicates that the MW driving does not generate  transitions between the $|1\rangle$ and $|-1\rangle$. 

For DQM we aim to induce rotations between the $|1\rangle$ and $|-1\rangle$ states by, e.g., a three-pulse-sequence involving the $|0\rangle$ state. In particular,  in a rotating frame with respect to (w.r.t.)  $DS_z^2 + |\gamma_e| B_z S_z$, Hamiltonian~(\ref{rotations}) is 
\begin{equation}\label{control}
H = \frac{\Omega(t)}{2} (e^{-i\varphi} |\pm 1\rangle\langle 0| + e^{i\varphi} |0\rangle\langle \pm1|),
\end{equation}
where the Rabi frequency $\Omega(t) = \gamma_e B^x_d(t)$, and the presence of $|\pm1\rangle$  is selected by tuning $\omega_{\rm MW} = D \pm |\gamma_e| B_z$ respectively. 
We denote the propagator (in the following we call it pulse) associated to Hamiltonian~(\ref{control}) as
\begin{equation}\label{simplepulse}
U_{[2\phi_{\rm f} ,\varphi]}^{\pm1}(t) = \exp\bigg[-i\phi(t) (e^{-i\varphi} |\pm 1\rangle\langle 0| + e^{i\varphi} |0\rangle\langle \pm1|)  \bigg],
\end{equation}
where  $\phi(t) = \int_{t_0}^{t} \frac{\Omega(s)}{2} ds$ with $t>t_0$, and $\phi_{\rm f}$ is the final achieved phase after the application of the MW driving. Then, the three-pulse-sequence $\tilde{U}^{[+1,-1,+1]}_{[\pi,0]}=U_{[\pi,0]}^{+1}U_{[\pi,0]}^{-1}U_{[\pi,0]}^{+1}$ is equivalent to an effective $\pi$ pulse on the $S_z$ operator. This is 
\begin{equation}\label{threepulse}
\bigg(\tilde{U}_{[\pi,0]}^{[+1,-1,+1]}\bigg)^{\dag} \ S_z \ \tilde{U}_{[\pi,0]}^{[+1,-1,+1]}  = -S_z.
\end{equation}
Note that the previous result persists if one uses the alternative $\tilde{U}_{[\pi,\pi/2]}^{[-1,+1,-1]}$ three-pulse-sequence. See Suplemental Material~\cite{Supplemental} for an explanation of the three-pulse-sequence effect on the S=1 spin manifold~\cite{Allard01} of the NV.

When a nuclear spin cluster is introduced in the formalism, Eq.~(\ref{rotations}) is completed to  
\begin{eqnarray}\label{pmmodel}
H =&& DS_z^2 + |\gamma_e| B_z S_z + \sqrt{2}\gamma_e B^x_d \cos{(\omega_{\rm MW} t + \varphi)} S_x\nonumber\\
&-&\sum_j \gamma_n B_z I_j^z + S_z\sum_j \vec{A}_j \cdot \vec{I}_j. 
\end{eqnarray}
Here, $\gamma_n$ is the nuclear gyromagnetic ratio, and $\vec{A}_j$ is the hyperfine vector that mediates the interaction between an NV and the $j$th nucleus~\cite{Maze08}.  In a rotating frame w.r.t. $DS_z^2 + |\gamma_e| B_z S_z$,  Hamiltonian~(\ref{pmmodel}) reads
\begin{eqnarray}\label{forsimulation}
H =&-&\sum_j \gamma_n B_z I_j^z + S_z\sum_j \vec{A}_j \cdot \vec{I}_j \nonumber\\
    &+&  \frac{\Omega(t)}{2} (e^{-i\varphi} |\pm 1\rangle\langle 0| + e^{i\varphi} |0\rangle\langle \pm1|).
\end{eqnarray}
We want to remark that we will use this Hamiltonian as the starting point for the following numerical simulations without doing any further assumption.

Typically, an NV spin qubit is selected between the  $|0\rangle$ and one of the $|\pm1\rangle$ hyperfine states. This is achieved via the  next transformation  (note we take $|0\rangle$ and $|+1\rangle$ as the NV qubit states) 
\begin{eqnarray}\label{simplification}
&&S_z\sum_j \vec{A}_j \cdot \vec{I}_j = \bigg(|1\rangle\langle 1| -  |-1\rangle\langle -1| \bigg) \sum_j \vec{A}_j \cdot \vec{I}_j\nonumber\\
&&=\frac{|1\rangle\langle 1| -  |0\rangle\langle 0|}{2} + \frac{|1\rangle\langle 1| + |0\rangle\langle 0|}{2} \sum_j \vec{A}_j \cdot \vec{I}_j \nonumber\\
&& \ \ - |-1\rangle\langle -1| \sum_j \vec{A}_j \cdot \vec{I}_j.
\end{eqnarray}

Then, if the MW driving in Eq.~(\ref{control}) does not induce transitions to  $|-1\rangle$ and the initial NV quantum state does not have  a $|-1\rangle$ component, the last line in Eq.~(\ref{simplification}) can be safely removed. Consequently, one can establish the equivalence 
\begin{equation}\label{equivalence}
S_z\sum_j \vec{A}_j \cdot \vec{I}_j \equiv  \frac{\sigma_z}{2} \sum_j \vec{A}_j \cdot \vec{I}_j  + \frac{I}{2} \sum_j \vec{A}_j \cdot \vec{I}_j,
\end{equation}
with  $\sigma_z = |1\rangle\langle 1| -  |0\rangle\langle 0|$, and $I = |1\rangle\langle 1| + |0\rangle\langle 0|$. On the one hand, this approach known as single quantum magnetometry (SQM) presents the advantage of having easily implementable $\pi$ pulses owing to the presence of a direct transition between the NV qubit states $|0\rangle$ and $|1\rangle$. On the other hand, the NV-nuclei coupling is reduced by a factor of $2$, see first term at the right hand side of Eq.~(\ref{equivalence}). From a different perspective, in SQM protocols the NV induced magnetic field gradient ($\frac{I}{2} \sum_j \vec{A}_j \cdot \vec{I}_j$) is useful to individually control $^{13}$C spins in the diamond lattice~\cite{Abobeih19} for, e.g., the implementation of quantum algorithms~\cite{Rosskopf17,Casanova17}. However, as we will demonstrate later, the term ($\frac{I}{2} \sum_j \vec{A}_j \cdot \vec{I}_j$) introduces an inhomogeneous broadening leading to misinterpretation in the obtained NMR spectrum. This seriously affects to the determination of important structural parameters such as the chemical shift. 
 
\begin{figure}[t]
\hspace{-0.4 cm}\includegraphics[width=0.96\columnwidth]{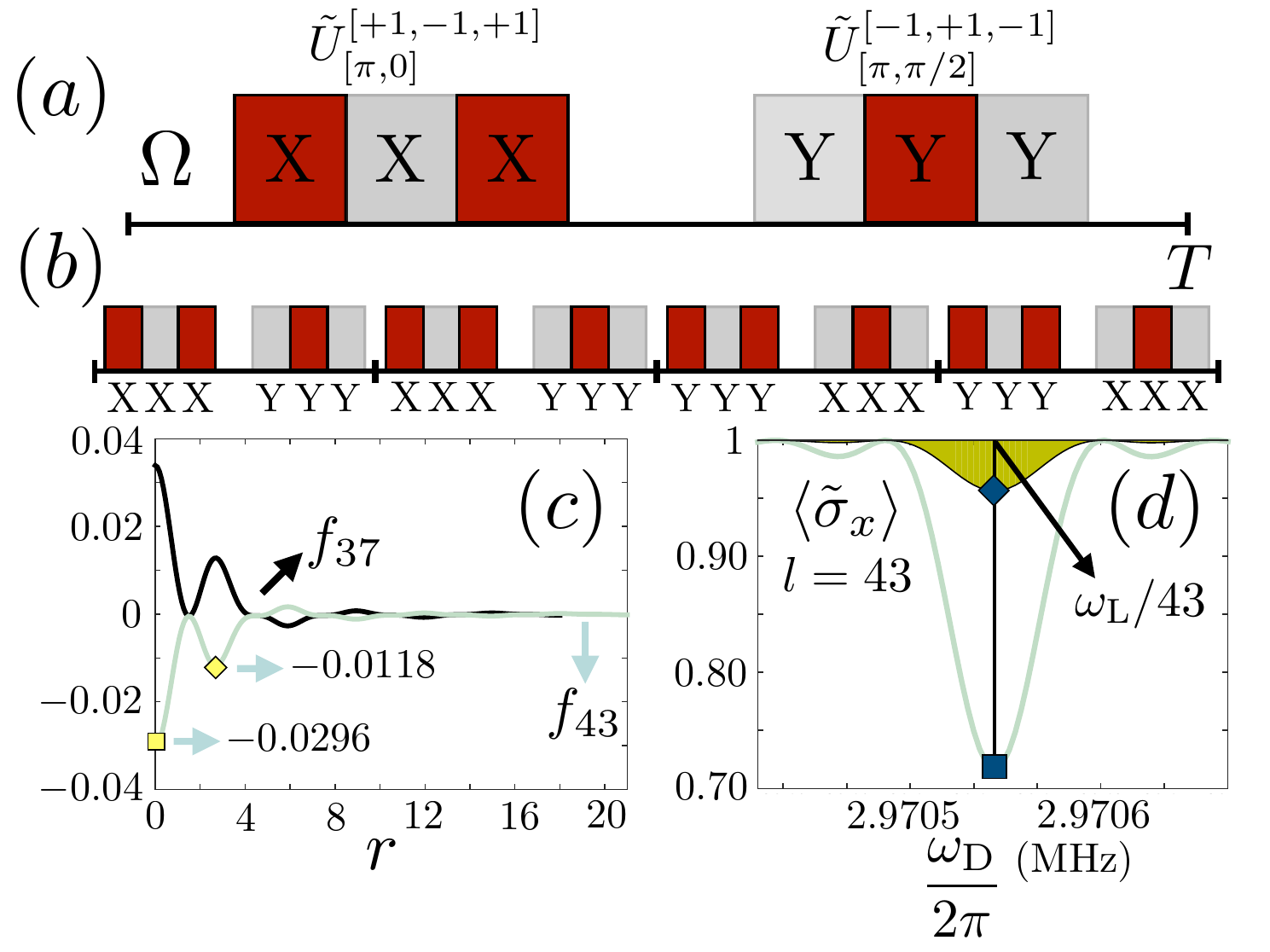}
\caption{(a)  Top-hat $\Omega$ leading to $\tilde{U}^{[+1,-1,+1]}_{[\pi,0]}$ and $\tilde{U}^{[-1,+1,-1]}_{[\pi,\pi/2]}$. (b) Pulse structure of the sequence we repeat in our numerical simulations. (c) $f_{37,43}(r)$ (dark and clear curves, respectively).  Yellow square and  diamond marked the values of $f_{43}(r)$ we use in the next figure. (d) Computed signal for instantaneous $\pi$ pulses (clear green curve) and for finite-width pulses (curve over dark-green area). Blue square and diamond mark the theoretically expected signal.}\label{scheme}
\end{figure}

We deal with the main disadvantage of DQM schemes (i.e. the absence of a direct transition between $|1\rangle$ and $|-1\rangle$) with the three-pulse-sequences $\tilde{U}^{[+1,-1,+1]}_{[\pi,0]}$ and $\tilde{U}_{[\pi,\pi/2]}^{[-1,+1,-1]}$. Nevertheless, we find that the performance of DQM schemes at the optimal situation of large $B_z$ gets challenging. In particular, when a set of $\pi$ pulses is applied on the $S_z$ operator, Hamiltonian~(\ref{forsimulation}) (in a rotating frame w.r.t. the nuclear spin energy $ -\sum_j \gamma_n B_z I_j^z$ and to the MW driving $\frac{\Omega(t)}{2} (e^{-i\varphi} |\pm 1\rangle\langle 0| + e^{i\varphi} |0\rangle\langle \pm1|)$) reads

\begin{equation}\label{evolving}
H= F(t) \ S_z \sum_j \bigg[ A_j^x I_j^x \cos{(\omega_{\rm L} t)} +  A_j^y I_j^y \sin{(\omega_{\rm L} t)} + A_j^z I_j^z\bigg].
\end{equation}
Here, the Larmor frequency $\omega_{\rm L} = \gamma_n B_z$, and the NV-nuclei coupling constants are $A_{j}^{x,y} = |\vec{A}_{j}^{x,y}|$ where $\vec{A}_{j}^{x} = \vec{A}_{j} -  (\vec{A}_{j}\cdot \hat{z}) \ \hat{z}$, $\vec{A}_{j}^{y} =   \hat{z}\times \vec{A}_{j}$, and $I_{x}^j = \vec{I}_j \cdot \hat x_j$,  $I_{y}^j = \vec{I}_j \cdot \hat{y}_j$ with $\hat{x}_j = \vec{A}_{j}^{x}/ A_{j}^{x}$
and $\hat{y}_j = \vec{A}_{j}^{y}/ A_{j}^{y}$. If we introduce a sequence of $\pi$ pulses ordered in an even manner, $F(t)$ can be written as a sum of even harmonic functions as $F(t) = \sum_k f_k \cos{(k \omega_{D} t)}$. Here $\omega_{\rm D} = \frac{2\pi}{T}$ with $T$ being the elemental period of the sequence, see Fig.~\ref{scheme} (a), and the Fourier coefficients $f_k = 2/T \int_0^T F(s)\cos{(k \omega_{D} s)} \ ds$. Now, under the resonance condition (for the $l$th harmonic)
\begin{equation}
l\omega_{D} = \omega_{\rm L},
\end{equation}
one finds the resonant NV-nuclei coupling Hamiltonian
\begin{equation}\label{onresonance}
H = \frac{f_l}{2} S_z \sum_j A_j^{x} I_j^{x},
\end{equation}
which reveals that the interaction strength of the NV with the nuclei depends on the value of $f_l$. For sequences with equally spaced instantaneous $\pi$ pulses, one can calculate that $f_l = (-1)^{\frac{l-1}{2}} \frac{4}{\pi l}$. However, at large $B_z$ the assumption of instantaneous pulses imply the transfer of a large amount of power to the sample. More specifically, in the typical case of top-hat $\pi$ pulses (i.e. pulses generated with $H = \frac{\Omega}{2} (e^{-i\varphi} |\pm~1\rangle\langle 0| + e^{i\varphi} |0\rangle\langle \pm1|)$ where the Rabi frequency $\Omega$ is constant) it is needed $\Omega\gg \omega_{\rm L}$ to assure the instantaneous character of the $\pi$ pulses (note that at $B_z = 3$ T~\cite{Aslam17} the Larmor frequency is $\omega_{\rm L} \approx (2\pi)\times 127$ MHz). This is because the $\pi$ pulse time is $t_\pi = \frac{\pi}{\Omega}$, then the condition $\frac{1}{2\Omega} \ll \frac{1}{\omega_{\rm L}}$ must hold to certify that Hamiltonian~(\ref{evolving}) does not induce any evolution in the system during the $\pi$ pulse duration.  A failure of this condition is adverse for the NV nuclei coupling, since the $f_l$ coefficient rapidly drops to zero. More specifically, one can calculate that, for odd $k$, and top-hat $\pi$ pulses 
\begin{equation}\label{decay}
f_l(r) =  \frac{(-1)^{\frac{l+1}{2}} 36 \cos^3{(\pi r)}}{\pi l(-9 + 36 r^2)},
\end{equation}
where $r$ determines the relation of $t_{\pi}$ with the nuclear Larmor period $\frac{2\pi}{\omega_{\rm L}}$ through $ t_{\pi} = \frac{2\pi r}{\omega_{\rm L}}$, see~\cite{Supplemental}. Equation~(\ref{decay}) predicts a decay $\sim \frac{1}{r^2}$,  while in the limit $r\rightarrow 0$ we recover the situation with instantaneous pulses, i.e.  $f_l = (-1)^{\frac{l-1}{2}} \frac{4}{\pi l}$. 
In addition, one can find that $f_l = 0$ for even $l$. Finally, we point out that the issue of the decreasing $f_l$  was previously shown in SQM~\cite{Casanova18} while here we observe it persists for DQM schemes.

In Fig.~\ref{scheme} (a) there is a scheme of the top-hat pulses leading to the three-pulse-sequences  $\tilde{U}^{[+1,-1,+1]}_{[\pi,0]}$ and $\tilde{U}^{[-1,+1,-1]}_{[\pi,\pi/2]}$. In particular, dark-red pulses imply a $\pi$ pulse between $|0\rangle$ and $|1\rangle$, while clear-grey between 
$|0\rangle$ and $|-1\rangle$. Furthermore, an X (Y) pulse corresponds to select $\varphi=0$ ($\pi/2$) in Eq.~(\ref{control}). In Fig.~\ref{scheme} (b) 
we sketch the pulse sequence we will use in our numerical simulations for our DQM schemes. In Fig.~\ref{scheme} (c) we show the value of  $f_l(r)$ (for $l=37$ and $43$). We observe that $f_{37,43}(r)$ decreases with $r$, leading to weaker values of the effective NV-nuclei coupling. This effect is confirmed in Fig.~\ref{scheme} (d) where we simulate Eq.~(\ref{forsimulation}) and compute the signal $\langle \tilde{\sigma}_x\rangle$ (with $\tilde{\sigma}_x  =  |1\rangle\langle-1| +  |-1\rangle\langle 1|$) for different values of $\omega_{\rm D}$. For the sake of simplicity, we consider one NV and a single H nucleus such that the hyperfine vector is $\vec{A} = (2\pi)\times[7.39,29.90,-4.61]$ kHz. Firstly, we simulate a situation with instantaneous pulses (clear-green curve) in an interval of $\omega_{\rm D}$ that includes the resonance condition $43\omega_{\rm D} = \omega_{\rm L}$. In this latter case (i.e. on resonance) the NV-nucleus dynamics is governed by the Hamiltonian $H=\frac{f_{43}}{2}S_z A^x I^x$ with $f_{43} =- \frac{4}{43\pi} = -0.0296$, see Eq.~(\ref{onresonance}). Note this is the value of $f_{43}$ that corresponds to instantaneous $\pi$ pulses, and it is marked in Fig.~\ref{scheme} (c) with a yellow square. The clear-green curve appears after repeating 200 times the sequence in Fig.~\ref{scheme} (b), which implies the delivery of 4600 $\pi$ pulses on the $|0\rangle \leftrightarrow |\pm 1\rangle$ NV spin states (final sequence time is $\approx$ 0.194 ms). The curve over the dark-green area has been computed by applying finite-width top-hat $\pi$ pulses such that $t_{\pi} \approx 7$ ns. This implies $\pi$ pulses generated with $\Omega \approx (2\pi)\times71$ MHz. The associated $f_{31}\approx -0.0158$ coefficient is marked in Fig.~\ref{scheme} (c) with a yellow diamond. In Fig.~\ref{scheme} (d) we compare the numerically obtained signal with the expected theoretical results according to the formula $\langle \tilde{\sigma}_x\rangle= \cos{(\frac{f_{43}}{2} A^x t_f)}$, see~\cite{Supplemental}. The blue square is the signal associated to $f_{43} =  -0.0296$ (instantaneous pulses) and blue diamond to  $f_{43} =  -0.0118$ (finite-width pulses). It can be observed a clear convergence between numerics and the theoretically expected signal, which confirms the presence of a severe signal reduction with realistic finite-width pulses. 

\section{Demonstration of the method} To overcome this serious problem that limits the applicability of DQM schemes at the important regime of large $B_z$, we introduce a tailored Rabi frequency $\Omega(t)$ to generate the three-pulse-sequences $\tilde{U}^{[+1,-1,+1]}_{[\pi,0]}$ and $\tilde{U}^{[-1,+1,-1]}_{[\pi,\pi/2]}$. In this manner we will recover the value $f_l = (-1)^{\frac{l-1}{2}} \frac{4}{\pi l}$ corresponding to instantaneous pulses, with realistic finite-width $\pi$ pulses that extends over several Larmor periods. In addition, and as we  show in the next figure, these extended $\pi$ pulses can be generated with only moderate MW power. At this point we want to comment that there are different manners of designing $\Omega(t)$, while we select the one described in the Supplemental Material~\cite{Supplemental}.

\begin{figure}[t]
\hspace{-0. cm}\includegraphics[width=1.0\columnwidth]{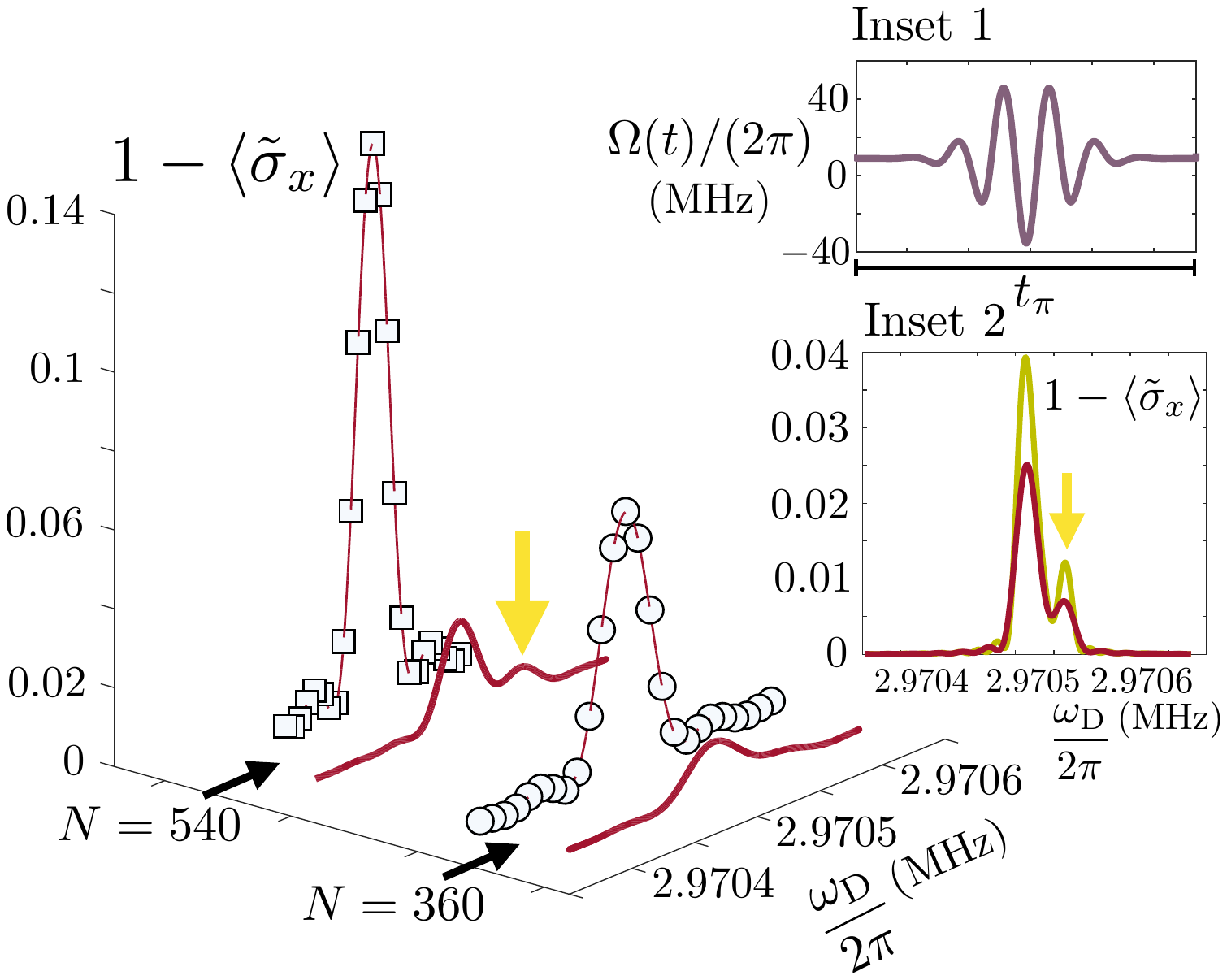}
\caption{Solid-red curves (wide curves) corresponds to ideal SQM schemes (i.e.  these signals appear after the applying instantaneous $\pi$ pulses). Thin-red lines are the signals appearing with DQM with instantaneous $\pi$ pulses, while circles and squares have been obtained with the modulated $\Omega(t)$ in the Inset 1.  Inset 2, emergence of a secondary peak. Note this does not occur in the case of our DQM schemes with tailored $\Omega(t)$.}
\label{tailoredF}
\end{figure}
 
In Fig.~\ref{tailoredF}  we compare the signals (for a better comparison we plot $1-\langle\tilde{\sigma}_x\rangle$) one can get by using SQM and DQM schemes in a 5-H spin cluster at an average distance from the NV sensor of $\approx 3.1$ nm. The hyperfine vectors of the nuclear cluster are $\vec{A}_1=(2\pi) \times[0.97,3.18,-4.14]$, $\vec{A}_2=(2\pi) \times[-2.09,2.69,0.44]$, $\vec{A}_3=(2\pi) \times[-1.84,3.18,-0.13]$, $\vec{A}_4=(2\pi) \times[-2.25,3.42,-0.69]$, and $\vec{A}_5=(2\pi) \times[-1.73,3.00,-1.22]$ kHz. All signals in Fig.~\ref{tailoredF} have been computed at a large $B_z = 3$ T. The solid-red curves in Fig.~\ref{tailoredF} correspond to signals obtained with SQM  protocols that employ instantaneous $\pi$ pulses (i.e. non-realistic) ordered as $({\rm XYXYYXYX})^N$ for $N=360, 540$. Note that for SQM, a single $\pi$ pulse is enough to flip the $\sigma_z$ operator in Eq.~(\ref{equivalence}). Thin-red-lines and overlapping circles and squares in Fig.~\ref{tailoredF} are the obtained signals for DQM schemes. In particular, thin-red-lines  assume instantaneous $\pi$ pulses following the ordering in Fig.~\ref{scheme} (b), while the overlapping circles and squares are obtained with our method involving a tailored $\Omega(t)$ that is shown in the Inset 1 (for the plot we repeated a number $N=360, 540$ the ideal sequence made of instantaneous pulses, and the realistic sequence with finite-width pulses). We stress that, this $\Omega(t)$ has been used to generate each $\pi$ pulse in the three-pulse-sequences  $\tilde{U}^{[+1,-1,+1]}_{[\pi,0]}$ and $\tilde{U}^{[-1,+1,-1]}_{[\pi,\pi/2]}$. Furthermore, as it can be clearly appreciated in Inset 1, the maximum value for the tailored Rabi frequency is $\approx (2\pi)\times 40$ MHz while $t_{\pi} \approx 0.16 \ \mu$s.
Then, even under the latter conditions of moderate MW power and long $\pi$ pulses that extend over several Larmor periods (recall that, for H nuclei at $B_z = 3$ T one Larmor period is $\approx 7.8$ ns) we reproduce the signal corresponding to an ideal DQM scheme  using instantaneous $\pi$ pulses (i.e. with infinite MW power). In addition, the simulations leading to circles and squares in Fig.~\ref{tailoredF} have been performed by considering an error of $1\%$ in the MW field drivings, as well as as an energy shift of $(2\pi)\times 20$ kHz on the NV spin transitions $|0\rangle \leftrightarrow |\pm 1\rangle$. In these conditions we observe no deviation with respect to the ideal case of instantaneous $\pi$ pulses (thin-red-lines) which certifies the robustness of our method.

Now we can better analyse the benefits of our method in nanoscale NMR. Firstly, in Fig.~\ref{tailoredF}, it gets clear that with DQM schemes we obtain signals with a larger contrast than those obtained with SQM protocols. Secondly, the effect of the inhomogeneous broadening in SQM can be already appreciated in Fig.~\ref{tailoredF} with the emergence of a secondary peak (marked with a yellow arrow) in the case of $N=540$. To better visualise it, in the Inset 2 we have plotted the SQM signals for $N=540$ (solid-red) and for $N=720$ (clear-green). Here it can be clearly seen the secondary peak that appears as a result of the magnetic field gradient ($\frac{I}{2} \sum_j \vec{A}_j \cdot \vec{I}_j$) introduced by SQM protocols. It is important to remark that these unwanted resonances induce ambiguities in the physical interpretation of the spectrum, since any secondary peak can be understood as arising from, e.g., a chemical shift  that deviates the resonance frequencies of the nuclei in the sample. On the contrary, as it is demonstrated in Fig.~\ref{tailoredF}, our DQM schemes produce clear peaks at the natural resonance frequency of the nuclei (i.e. when $43\omega_{D} = \omega_{\rm L}$) which confirms that our method does not contaminate the nuclear cluster with magnetic field gradients. Finally, owing to the introduced design for $\Omega(t)$, our DQM methods can operate in the regime of large $B_z$ fields where chemical shifts get enhanced~\cite{Levitt08}.  

\section{Conclusions} We presented a method for DQM at large static magnetic fields. This is a regime where important parameters, such as the chemical shift that encodes structural information, get enhanced. Via the introduction of appropriately tuned $\Omega(t)$ we get larger contrasts in the nanoscale NMR signal. In addition, our method avoids spectral broadenings induced by the magnetic field gradients that appear in commonly used SQM protocols. This is of special importance for the measurements of natural frequency deviations in the nuclear spins of the sample. Our method is general, since it can be adapted to any DD sequence while it is applicable in different quantum sensors. 

\begin{acknowledgements}
Authors acknowledge financial support from Spanish Government PGC2018-095113-B-I00 (MCIU/AEI/FEDER, UE), Basque Government IT986-16, as well as from QMiCS (820505) and OpenSuperQ (820363) of the EU Flagship on Quantum Technologies, and to the EU FET Open Grant Quromorphic (828826). J.C. acknowledges support by the Juan de la Cierva grant IJCI-2016-29681, and to the UPV/EHU grant EHUrOPE. I. A. acknowledges support to the Basque Government PhD grant PRE-2015-1-0394. 
\end{acknowledgements}

\pagebreak
\widetext
\begin{center}
\textbf{ \large Supplemental Material: \\ Double Quantum Magnetometry at Large Static Magnetic Fields}
\end{center}
\tableofcontents

\setcounter{equation}{0} \setcounter{figure}{0} \setcounter{table}{0}
\setcounter{section}{0}
\setcounter{page}{1} \makeatletter \global\long\def\theequation{S\arabic{equation}}
 \global\long\def\thefigure{S\arabic{figure}}
 \global\long\def\bibnumfmt#1{[S#1]}
 \global\long\def\citenumfont#1{S#1}

\starttocentries
\section{Three-Pulse-sequences}\label{Sbigsec}

Here, we show in detail the achievement of the modulation function of the $S_z$ operator involved in the three-pulse-sequences $\tilde{U}^{[+1,-1,+1]}_{[\pi,0]}$ and $\tilde{U}_{[\pi,\pi/2]}^{[-1,+1,-1]}$.

\subsection{First three-pulse-sequence $\tilde{U}^{[+1,-1,+1]}_{[\pi,0]}$}

One can demonstrate that a $2\phi$ pulse of the kind  ${\rm exp}[{\pm i  \phi (e^{-i\varphi} |\pm 1\rangle\langle 0| + e^{i\varphi}  |0\rangle\langle \pm 1|)}]$ on the $S=1$ manifold of the NV center is 
\begin{eqnarray}\label{xpulse}
U^{\pm1}_{[\pm 2\phi,\varphi]} &=& {\rm exp}[{\pm i  \phi (e^{-i\varphi} |\pm 1\rangle\langle 0| + e^{i\varphi}  |0\rangle\langle \pm 1|)}]\nonumber\\ 
&=& I + (\cos{\phi} -1)  (|\pm 1\rangle\langle \pm 1| + | 0\rangle\langle  0|) 
\pm i \sin{\phi}  ( e^{-i\varphi} |\pm 1\rangle\langle 0| + e^{i\varphi}  |0\rangle\langle \pm 1|),
\end{eqnarray}
where $I =  | 1\rangle\langle  1| + | 0\rangle\langle  0| + |- 1\rangle\langle - 1|$.

\subsubsection{First pulse}
In this manner, the application of a $2\phi$ pulse over the X axis  (i.e. $\varphi = 0$) and acting, e.g., on the $\bigg\{ |1\rangle, |0\rangle \bigg\}$ spin manifold transforms  the $S_z$ operator as 
\begin{eqnarray}
&&{\rm exp}[{+ i  \phi ( | 1\rangle\langle 0| + |0\rangle\langle 1|)}] S_z {\rm exp}[{- i  \phi ( | 1\rangle\langle 0| + |0\rangle\langle 1|)}] = \nonumber\\
&&\cos{\phi}^2 |1\rangle\langle 1| + \sin{\phi}^2 |0\rangle\langle 0| - |-1\rangle \langle -1 | -i\cos\phi \sin\phi ( | 1\rangle\langle 0| - |0\rangle\langle 1|).
\end{eqnarray}

The basis of an $S=1$  Hilbert space has a dimension $d=9$. Then, by using a similar  basis (note we restrict to only the diagonal terms) for $S=1$ than the one proposed one in~\cite{Allard01} we get
\begin{eqnarray}\label{Stransformation}
|1\rangle\langle1| &=& \frac{1}{3} I +  \frac{1}{2} S_z +  \frac{1}{6} G,\nonumber\\
|0\rangle\langle0| &=& \frac{1}{3} I - \frac{1}{3} G,\nonumber\\
|-1\rangle\langle-1| &=& \frac{1}{3} I -  \frac{1}{2} S_z +  \frac{1}{6} G,
\end{eqnarray}
where 
\begin{eqnarray}
I &=&  | 1\rangle\langle  1| + | 0\rangle\langle  0| + |- 1\rangle\langle - 1|,\nonumber\\
S_z &=&  | 1\rangle\langle  1| - |- 1\rangle\langle - 1|, \nonumber\\
G &=&  | 1\rangle\langle  1| -2 | 0\rangle\langle  0| + |- 1\rangle\langle - 1|.
\end{eqnarray}
For a final $\phi = \pi/2$ (i.e. we applied the $U^{+1}_{[\pi,0]}$ operator, this is a $\pi$ pulse over the X axis in the $\bigg\{ |1\rangle, |0\rangle \bigg\}$ spin manifold) $S_z$ transforms to $|0\rangle\langle 0| - |-1\rangle \langle -1 | $, while during the application of the X pulse $S_z$ evolves as 
$\frac{1}{2} (\cos^2\phi + 1)$. 

Then, during this first pulse we have that $S_z$ transforms as
\begin{equation}\label{Sfirstpulse}
S_z \longrightarrow \frac{1}{2} (\cos^2\phi + 1) \ S_z - \frac{1}{2}  (\sin^2\phi)  \ G. 
\end{equation}
While for the final value of  $\phi = \pi/2$ we find that $S_z$ has changed as  $ S_z \longrightarrow \frac{1}{2} \ S_z - \frac{1}{2} \ G = |0\rangle\langle 0| - |-1\rangle \langle -1 |$.
\subsubsection{Second pulse}

The next step is to interchange the $|0\rangle$ and $|-1\rangle$ spin components. This is achieved by applying the $U^{-1,0}_{\pm 2\phi,0}$ pulse operator for a final value of $\phi=\pi/2$.  More specifically, this is 
\begin{equation}
U^{-1,0}_{ 2\phi,0} \bigg( |0\rangle\langle 0| - |-1\rangle \langle -1 | \bigg) U^{-1,0}_{ -2\phi,0} = (\cos^2\phi - \sin^2\phi)  \bigg( |0\rangle\langle 0| - |-1\rangle \langle -1 | \bigg) -2i\sin\phi \cos \phi \ \bigg(|0\rangle \langle -1| - |-1\rangle \langle 0|\bigg). 
\end{equation}
Then, by using again the transformation in Eq.~(\ref{Stransformation}) one can find
\begin{equation}
\frac{1}{2} \ S_z - \frac{1}{2} \ G = |0\rangle\langle 0| - |-1\rangle \langle -1 | \longrightarrow \frac{1}{2} (\cos^2\phi - \sin^2\phi) (S_z - G_z).
\end{equation}
Again, for the final value of $\phi=\pi/2$ the final state is $\frac{1}{2} (G_z - S_z) = -  \bigg(|0\rangle\langle 0| - |-1\rangle \langle -1 |\bigg) $

\subsubsection{Third pulse}

Finally, we apply a last pulse  defined by $U^{1,0}_{ 2\phi,0}$. This is
\begin{eqnarray}
&&U^{1,0}_{ 2\phi,0} \bigg( -|0\rangle\langle 0| + |-1\rangle \langle -1 | \bigg) U^{1,0}_{ -2\phi,0} =\nonumber\\
&&-\cos^2\phi |0\rangle\langle 0| -\sin^2\phi |1\rangle\langle 1| + |-1\rangle\langle -1| + i\cos\phi \sin\phi ( |0\rangle\langle 1|- |1\rangle\langle 0|).
\end{eqnarray}
This implies that, during this last pulse, the $\frac{1}{2} (G_z - S_z) = -  \bigg(|0\rangle\langle 0| - |-1\rangle \langle -1 |\bigg)$ operator changes as  
\begin{equation}
\frac{1}{2} (G_z - S_z) = -  \bigg(|0\rangle\langle 0| - |-1\rangle \langle -1 |\bigg) \longrightarrow -\frac{1}{2} [1 + \sin^2\phi] S_z +\frac{1}{2}  \cos^2\phi \ G
\end{equation}

Summarising, the evolution of the $S_z$ operator during each step of the three-pulse-sequence  $\tilde{U}^{[+1,-1,+1]}_{[\pi,0]}=U_{[\pi,0]}^{+1}U_{[\pi,0]}^{-1}U_{[\pi,0]}^{+1}$   is
\begin{eqnarray}
&{\rm First \ pulse: \ \ }  \frac{1}{2} (\cos^2\phi + 1) \ S_z - \frac{1}{2}  (\sin^2\phi)  \ G  \nonumber\\
&{\rm Second \ pulse: \ \ }\frac{1}{2} (\cos^2\phi - \sin^2\phi) (S_z - G_z) \nonumber\\
&{\rm \ Third \ pulse: \ \ }-\frac{1}{2} [1 + \sin^2\phi] S_z +\frac{1}{2}  \cos^2\phi \ G
\end{eqnarray}

\subsection{Second three-pulse-sequence $\tilde{U}_{[\pi,\pi/2]}^{[-1,+1,-1]}$}
Here we will study the second three-pulse-sequence $\tilde{U}_{[\pi,\pi/2]}^{[-1,+1,-1]}$ that would complete one period of the $S_z$ operator. \subsubsection{First pulse}
We apply the $U_{[\phi,\pi/2]}^{-1}$ pulse to the $-S_z$ operator (note this is the operator we get after using the previously explained $\tilde{U}_{[\pi, 0]}^{[+1,-1,+1]}$ on the initial $S_z$) and find 
\begin{eqnarray}
&&{\rm exp}[{+ i  \phi ( e^{-i\frac{\pi}{2}}| -1\rangle\langle 0| + e^{i\frac{\pi}{2}}|0\rangle\langle -1|)}] (- S_z) {\rm exp}[{- i  \phi (  e^{-i\frac{\pi}{2}}| -1\rangle\langle 0| +  e^{i\frac{\pi}{2}}|0\rangle\langle -1|)}] = \nonumber\\
&&\cos{\phi}^2 |-1\rangle\langle -1| + \sin{\phi}^2 |0\rangle\langle 0| - |1\rangle \langle 1 | -i\cos\phi \sin\phi (  e^{-i\frac{\pi}{2}} |- 1\rangle\langle 0| -  e^{i\frac{\pi}{2}} |0\rangle\langle -1|).
\end{eqnarray}
Then, we have that $-S_z$ changes as 
\begin{equation}
-S_z \longrightarrow  - \frac{1}{2} (\cos^2\phi + 1) \ S_z - \frac{1}{2}  (\sin^2\phi)  \ G.
\end{equation}
For the final value of $\phi = \frac{\pi}{2}$ we have $-S_z \longrightarrow  -\frac{1}{2} (S_z + G) = - (|1 \rangle \langle 1| - |0 \rangle \langle 0|)$
\subsubsection{Second pulse}
Now we apply the $U_{[\phi,\pi/2]}^{+1}$ pulse and get
\begin{eqnarray}
&&{\rm exp}[{+ i  \phi ( e^{-i\frac{\pi}{2}}| 1\rangle\langle 0| + e^{i\frac{\pi}{2}}|0\rangle\langle 1|)}] \ (|0 \rangle \langle 0| - |1 \rangle \langle 1|)) \ {\rm exp}[{- i  \phi (  e^{-i\frac{\pi}{2}}| 1\rangle\langle 0| +  e^{i\frac{\pi}{2}}|0\rangle\langle 1|)}] = \nonumber\\
&&(\cos{\phi}^2 - \sin{\phi}^2) \bigg(|0\rangle\langle 0| - |1\rangle \langle 1 |\bigg) -2i\cos\phi \sin\phi \ \bigg(  e^{i\frac{\pi}{2}} |0\rangle\langle 1| -  e^{-i\frac{\pi}{2}} |1\rangle\langle 0|\bigg).
\end{eqnarray}
Then, we have the following transformation 
\begin{equation}
-\frac{1}{2} (S_z + G)  \longrightarrow -\frac{1}{2} (\cos{\phi}^2 - \sin{\phi}^2) (S_z + G),
\end{equation}
while for the final value of $\phi = \pi/2$ we have $-\frac{1}{2} (S_z + G) =  (|1\rangle\langle 1| - |0\rangle\langle 0|)$

\subsubsection{Third pulse}
We finally use the $U_{[\phi,\pi/2]}^{-1}$ pulse to complete the sequence. This leads to
\begin{eqnarray}
&&{\rm exp}[{+ i  \phi ( e^{-i\frac{\pi}{2}}| -1\rangle\langle 0| + e^{i\frac{\pi}{2}}|0\rangle\langle -1|)}] (|1\rangle\langle 1| - |0\rangle\langle 0|) {\rm exp}[{- i  \phi (  e^{-i\frac{\pi}{2}}| -1\rangle\langle 0| +  e^{i\frac{\pi}{2}}|0\rangle\langle -1|)}] = \nonumber\\
&&-\cos{\phi}^2 |0\rangle\langle 0| - \sin{\phi}^2 |-1\rangle\langle -1| + |1\rangle \langle 1 | -i\cos\phi \sin\phi (  e^{-i\frac{\pi}{2}} |- 1\rangle\langle 0| -  e^{i\frac{\pi}{2}} |0\rangle\langle -1|).
\end{eqnarray}
Then, we have the transformation 
\begin{equation}
-\frac{1}{2} (S_z + G) =  (|1\rangle\langle 1| - |0\rangle\langle 0|) = \frac{1}{2} [1 + \sin^2\phi] S_z +\frac{1}{2}  \cos^2\phi \ G.
\end{equation}

Then, for a final value $\phi=\pi/2$ we recover the $S_z$ operator.

Summarising, the evolution of the $-S_z$ operator during the three-pulse-sequence $\tilde{U}_{[\pi,\pi/2]}^{[-1,+1,-1]}$ is
\begin{eqnarray}
&{\rm First \ pulse: \ \ }  -\frac{1}{2} (\cos^2\phi + 1) \ S_z - \frac{1}{2}  (\sin^2\phi)  \ G\nonumber\\
&{\rm Second  \ pulse: \ \ } -\frac{1}{2} (\cos^2\phi - \sin^2\phi) (S_z + G_z) \nonumber\\
&{\rm Third \ pulse: \ \ } +\frac{1}{2} [1 + \sin^2\phi] S_z +\frac{1}{2}  \cos^2\phi \ G
\end{eqnarray}

\section{Calculating the $f_l$ coefficient for top-hat pulses}\label{Stophat}
\begin{figure}[t]
\hspace{-0. cm}\includegraphics[width=0.7\columnwidth]{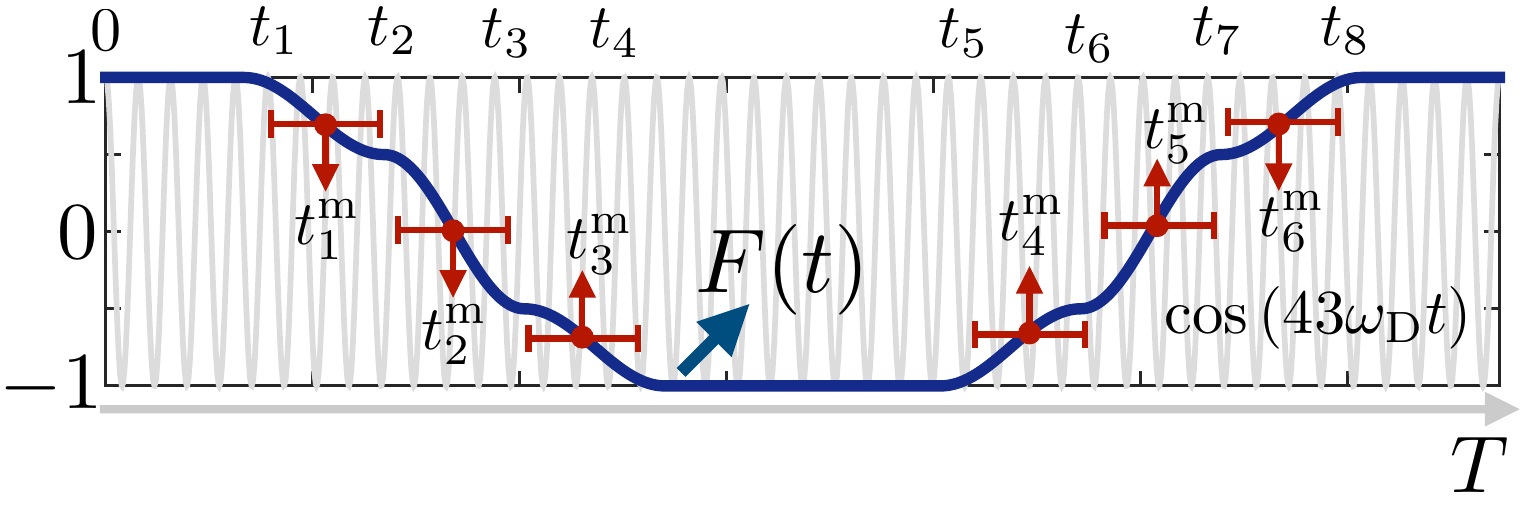}
\caption{Modulation function $F(t)$ (solid-blue) appearing after the application of the sequence in Eq.~(\ref{Ssequence}). For this plot, we have selected the $43$th harmonic as the target harmonic (see figure background in clear grey) since this is the one we use for the numerical simulations in the main text. For the sake of simplicity we assume $(t_2 - t_1) = (t_3 - t_2) =  (t_4 - t_3) = (t_6 - t_5) = (t_7 - t_6) = (t_8 - t_7) = t_\pi.$ In addition, in this plot it is marked the central point of each time interval where a $\pi$ pulse is displayed. This is, e.g., the firs top-hat $\pi$ pulse is displayed in the interval $(t_2-t_1)$ being $t_1^{\rm m}$ its center.}
\label{StailoredF}
\end{figure}

After the application of the following sequence to the $S_z$ operator
\begin{equation}\label{Ssequence}
\bigg(-~-~-\tilde{U}^{[-1,+1,-1]}_{[\pi,\pi/2]} -~-~-~-~-~- \tilde{U}^{[+1,-1,+1]}_{[\pi,0]}-~-~-\bigg), 
\end{equation}
note that with the symbol $``-~-~-$" we indicate the free evolution of the $S_z$ operator, one would get the modulation function $F(t)$ that appears in Fig.~\ref{StailoredF}. Also, we want to remark that the sequence in Eq.~(\ref{Ssequence}) appears when the top-hat pulses in Fig.~1 (a) of the main text is displayed. 

Then, taking into account the expressions for $F(t)$ calculated in the previous section and that we defined $t_{\pi} =\frac{2\pi r}{\omega_{\rm L}} $, one can easily find that  
\begin{equation}
f_{43} = \frac{2}{T}\int_0^T F(s) \cos{(43\omega_{\rm D} t)}ds = \frac{ 36 \cos^3{(\pi r)}}{\pi 43 (-9 + 36 r^2)},
\end{equation}
or, in general 
\begin{equation}
f_{l} = \frac{2}{T}\int_0^T F(s) \cos{(l\omega_{\rm D} t)}ds = \frac{(-1)^{\frac{l+1}{2}} 36 \cos^3{(\pi r)}}{\pi l (-9 + 36 r^2)}.
\end{equation}
\section{Theoretically expected signal}
For the resonant Hamiltonian $H=\frac{f_{43}}{2}S_z A^xI_x$ proposed in the main text, one can easily calculate that, for an initial state NV-	nucleus state $\rho_0= \frac{1}{2} (|1\rangle+|-1\rangle)(\langle1|+\langle-1|)\otimes \frac{1}{2}I$ (i.e. the initial NV state is the superposition $\frac{1}{\sqrt{2}} (|1\rangle+|-1\rangle)$, and the nucleus in a thermal state that we approximate to $\frac{1}{2} I_{2\times 2}$) the theoretically expected signal is 
\begin{equation}
\langle \tilde{\sigma}_x \rangle = {\rm Tr}\bigg( e^{-i \frac{f_{43}}{2}S_z A^xI_x t} \rho_0  e^{i \frac{f_{43}}{2}S_z A^xI_x t} \tilde{\sigma}_x\bigg) = \cos{\bigg(\frac{f_{43}}{2}A^x t\bigg)}.
\end{equation}

\section{Finding appropriate $\Omega(t)$ functions}
The shape of the $F(t)$ function that appears after the application of top-hat pulses can be seen in Fig.~\ref{StailoredF} and, as we have demonstrated in Sec.~\ref{Stophat}, it leads to a decreasing value for the $f_l$ coefficients. 

As we are introducing three $\pi$ pulses per flip of the $S_z$ operator, i.e. we need three $\pi$ pulses to generate the propagators $\tilde{U}^{[+1,-1,+1]}_{[\pi,0]}$ and $\tilde{U}^{[-1,+1,-1]}_{[\pi,\pi/2]}$ we  can always write a $f_l$ coefficient as (see Fig.~\ref{StailoredF})

\begin{eqnarray}\label{Slong}
f_l &=& \frac{2}{T}\bigg[  \int_0^{t_1} \cos{(l \omega_{\rm D} s)} ds + \int_{t_1}^{t_2} F(s) \cos{(l \omega_{\rm D} s)} ds +  \int_{t_2}^{t_3} F(s) \cos{(l \omega_{\rm D} s)} ds +  \int_{t_3}^{t_4} F(s) \cos{(l \omega_{\rm D} s)} ds\nonumber\\
&-& \int_{t_4}^{t_5} \cos{(l \omega_{\rm D} s)} ds + \int_{t_5}^{t_6} F(s) \cos{(l \omega_{\rm D} s)} ds +  \int_{t_6}^{t_7} F(s) \cos{(l \omega_{\rm D} s)} ds  + \int_{t_7}^{t_8} F(s) \cos{(l \omega_{\rm D} s)} ds \nonumber\\
&+&  \int_{t_8}^{T} \cos{(l \omega_{\rm D} s)} ds \bigg].
\end{eqnarray}

Then, if we find a $F(t)$ function such that it cancels the integral during the $\pi$ pulses, the above equation reduces to
\begin{equation}\label{Sreduced}
f_l = \frac{2}{T}\bigg[  \int_0^{t_1} \cos{(l \omega_{\rm D} s)} ds - \int_{t_4}^{t_5} \cos{(l \omega_{\rm D} s)} ds  +  \int_{t_8}^{T} \cos{(l \omega_{\rm D} s)} ds \bigg].
\end{equation}
Now, as it is standard in dynamical decoupling techniques we select $(t_1-0) = \frac{1}{2}(t_5-t_4) = (T-t_8)$. In addition, we assume that each $\pi$ pulse has the same duration, this is  $(t_2 - t_1) = (t_3 - t_2) =  (t_4 - t_3) = (t_6 - t_5) = (t_7 - t_6) = (t_8 - t_7) = t_\pi$, then it can be  calculated that Eq.~(\ref{Sreduced}) reads
\begin{equation}
f_l = \frac{4}{\pi l} \cos{\bigg(\pi \frac{3 t_\pi}{T/l}\bigg)}\sin{(\pi l/2)},
\end{equation} 
which takes the maximum value $|f_l| = \frac{4}{\pi l}$ (this is the value we target for our numerical simulations in the main text) when each modulated $\pi$ pulse takes an extension determined by
\begin{equation}
t_{\pi} = \frac{n}{3} T/l ,
\end{equation}
with $n$ being a natural number.

Now, we have to find a form for $F(t)$ such that it eliminates the integrals involving $\pi$ pulses in Eq.~(\ref{Slong}). Our solution correspond to introduce a $F(t)$ with two components. The first component is the shape that  $F(t)$ acquires in case of applying only top-hat $\pi$ pulses, the second one is an oscillating function such that it cancels the area of the first component. 

In particular, for the first $\pi$ pulse we propose that $F(t)$ takes the following form 
\begin{equation}\label{Sfirstmodulated}
F(t)= \frac{1}{2} \bigg[\cos^2\bigg(\frac{\pi(t -t_1)}{t_{\pi}}\bigg) + 1\bigg] - \alpha_1 \cos\bigg(\frac{2 \pi k}{T} t\bigg)
\exp[ -(t - t^{\rm m}_1)^2/(2\sigma_1^2)],
\end{equation}
with $t \in [t_2, t_1]$, $t_1^{\rm m}$ the central point of the interval $[t_2, t_1]$ and $\sigma_1$ the width (that one can freely adjust) of the Gaussian function in the second component of Eq.~(\ref{Sfirstmodulated}). Note that, as we have explained, the first component of Eq.~(\ref{Sfirstmodulated}) is $ \frac{1}{2} \bigg[\cos^2\bigg(\frac{\pi(t -t_1)}{t_{\pi}}\bigg) + 1\bigg]$, which is the aspect of $F(t)$ during a top-hat $\pi$ pulse, see Eq~(\ref{Sfirstpulse}).

Then the only missing parameter is the constant $\alpha_1$ that we will use to cancel the integral 
$\int_{t_1}^{t_2} F(s) \cos{(l \omega_{\rm D} s)} \  ds$, see Eq.~(\ref{Slong}). In this manner, the value of $\alpha_1$ is 
\begin{equation}
\alpha_1 = -\frac{\int_{t_1}^{t_2}  \frac{1}{2} \bigg[\cos^2\bigg(\frac{\pi(s -t_1)}{t_{\pi}}\bigg) + 1\bigg] \cos{(l \omega_{\rm D} s)} \ ds }{ \int_{t_1}^{t_2} \cos\bigg(\frac{2 \pi k}{T} t\bigg) \exp[ -(t - t^{\rm m}_1)^2/(2\sigma_1^2)]  \cos{(l \omega_{\rm D} s)}  \ ds}
\end{equation}

Finally, once all parameters are selected we can find the value of the Rabi frequency that generates the $F(t)$ in Eq.~(\ref{Sfirstmodulated}) as
\begin{equation}
\Omega(t) = \frac{\partial}{\partial_t} \arccos[F(t)].
\end{equation}
The previous equation can be understand by noting that the effect of the first driving $H = \frac{\Omega(t)}{2} (e^{-i\varphi} |+ 1\rangle\langle 0| + e^{i\varphi} |0\rangle\langle +1|)$ that acts over the $|0\rangle \leftrightarrow  |+1\rangle$ NV spin transition, is 
\begin{equation}
S_z \longrightarrow \frac{1}{2} \bigg(\cos^2 \int_{t_1}^t \frac{\Omega(s)}{2} \ ds + 1\bigg) \ S_z - \frac{1}{2}  (\sin^2\phi)  \ G. 
\end{equation}
Then, one only has to invert the equation 
\begin{equation}
F(t) = \frac{1}{2} \bigg(\cos^2 \int_{t_1}^t \frac{\Omega(s)}{2} \ ds + 1\bigg), 
\end{equation} 
to find $\Omega(t)$. If we do that, we get the modulated $\Omega(t)$ that appears in Fig.~2 of the main text.

The same strategy has to be applied to eliminate the integrals of all $\pi$ pulses in Eq.~(\ref{Sreduced}). Note that for each case (i.e. for each $\pi$ pulse leading to the three-pulse-sequences) the first component of $F(t)$ has a different aspect, but fortunately we have calculated all of them in Sec.~\ref{Sbigsec} of this Supplemental Material.

\end{document}